\documentclass[proceedings, preprint]{rmaa}

\usepackage{paralist}

\usepackage{psfrag,color}

\SetYear{2025}
\SetConfTitle{IX Reunión de Astronomía Dinámica en LatinoAmérica}

\title{The Astronomical Plate Digitization at SHAO} 

\author{
  Yong Yu,\altaffilmark{1}
  Meiting Yang,\altaffilmark{1}
  Zhengjun Shang,\altaffilmark{1}
  Liangliang Wang,\altaffilmark{1}
  Jing Yang,\altaffilmark{1}
  Zhenghong Tang,\altaffilmark{1}
  Jianhai Zhao,\altaffilmark{1}
  Massinissa Hadjara\altaffilmark{2,3}}

\altaffiltext{1}{Shanghai Astronomical Observatory, Chinese Academy Sciences, Shanghai 200030, China}
\altaffiltext{2}{Nanjing Institute of Astronomical Optics \& Technology, National Astronomical Observatories, Chinese Academy of Sciences, Nanjing 210042, China}
\altaffiltext{3}{Chinese High Angular Resolution Southern Astronomical Laboratory (CHARSAL)/Astro-Photonics Laboratory/Space and Planetary Exploration Laboratory (SPEL), Dept. of Electrical Engineering, FCFM, Universidad de Chile, Av. Tupper 2007, Santiago, Chile}

\shortauthor{Yu, Yang, Shang, Wang, Yang, Tang, Zhao \& Hadjara}
\shorttitle{SHAO's Plate Digitization}

\listofauthors{Y. Yu, M. Yang, Z. Tang, \& M. Hadjara}
\indexauthor{Yu, Y.}
\indexauthor{Yang, M.}
\indexauthor{Shang, Z.}
\indexauthor{Wang, L.}
\indexauthor{Yang, J.}
\indexauthor{Tang, Z.}
\indexauthor{Zhao, J.}
\indexauthor{Hadjara, M.}

\abstract{The digitization of historical astronomical plates is essential for preserving century-long observational data. This work presents the development and application of the specialized digitizers at the Shanghai Astronomical Observatory (SHAO), including technical details, international collaborations, and scientific applications on the plates.}

\resumen{}

\addkeyword{Astrometry}
\addkeyword{Methods: observational}
\addkeyword{instrumentation: detectors}
\addkeyword{techniques: image processing}

\begin{document}
\maketitle

\section{Introduction}

\begin{figure}[!t]
  \includegraphics[width=\columnwidth]{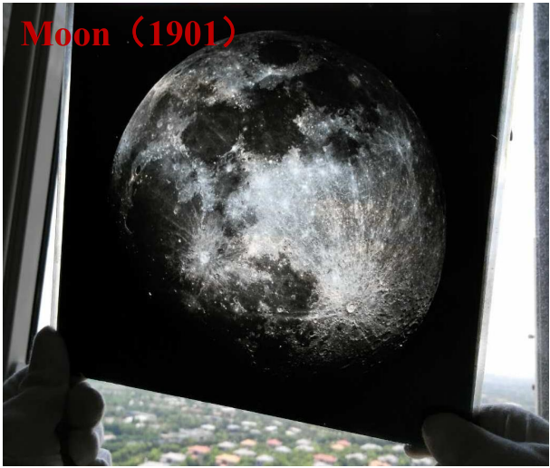}
  \includegraphics[width=\columnwidth]{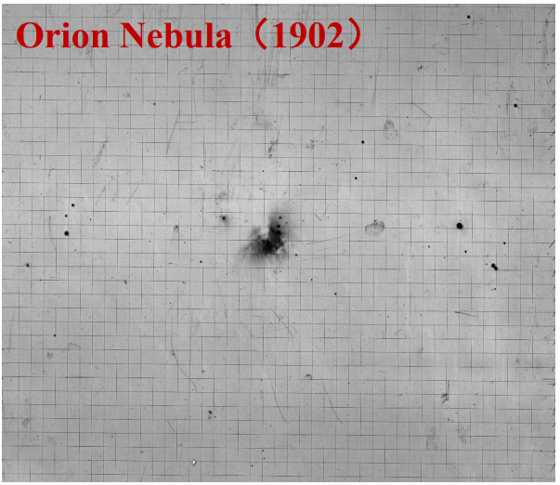}
  \caption{Examples of astronomical plates taken by SHAO.}
  \label{Fig1}
\end{figure}

Before the 1990s, astronomers used glass plates to record celestial bodies for over a century. These non-reproducible records are invaluable for studying long-term phenomena. However, degradation such as mildew and delamination has threatened their preservation. Samples of astronomical plates taken by SHAO are shown in Fig.~\ref{Fig1}.
To address this, and according to the International Astronomical Union (IAU) in Resolution B3 of its 24th General Assembly in 2000, SHAO established a dedicated Plate Laboratory in 2009, initiating a national effort to rescue and digitize Chinese astronomical plates. By the end of 2012, about 30,000 plates from all Chinese observatories were transferred to SHAO's Plates Digitization Laboratory (PD-Lab).

We began by performing performance tests on conventional commercial scanners. The results showed poor positioning repeatability in the scanning direction, poor illumination uniformity, and relatively high distortion. We then concluded that a specialized digitizer for astronomical plates was required for micrometer-level scanning accuracy.

\section{Development of Plate Digitizers at SHAO}

\subsection{First Digitizer (2013--2017)}

We naturally began by drawing inspiration from existing scientific digitizing machines. In 2011-2012, our PD-Lab team members visited the DASCH digitizer at Harvard College Observatory \citep{Simcoe2009} and the DAMIAN digitizer at the Royal Observatory of Belgium \citep{de Cuyper2006}, where we gained a wealth of knowledge about the operating principle and development of specialized digitizers.

\begin{figure}[!t]
  \includegraphics[width=\columnwidth]{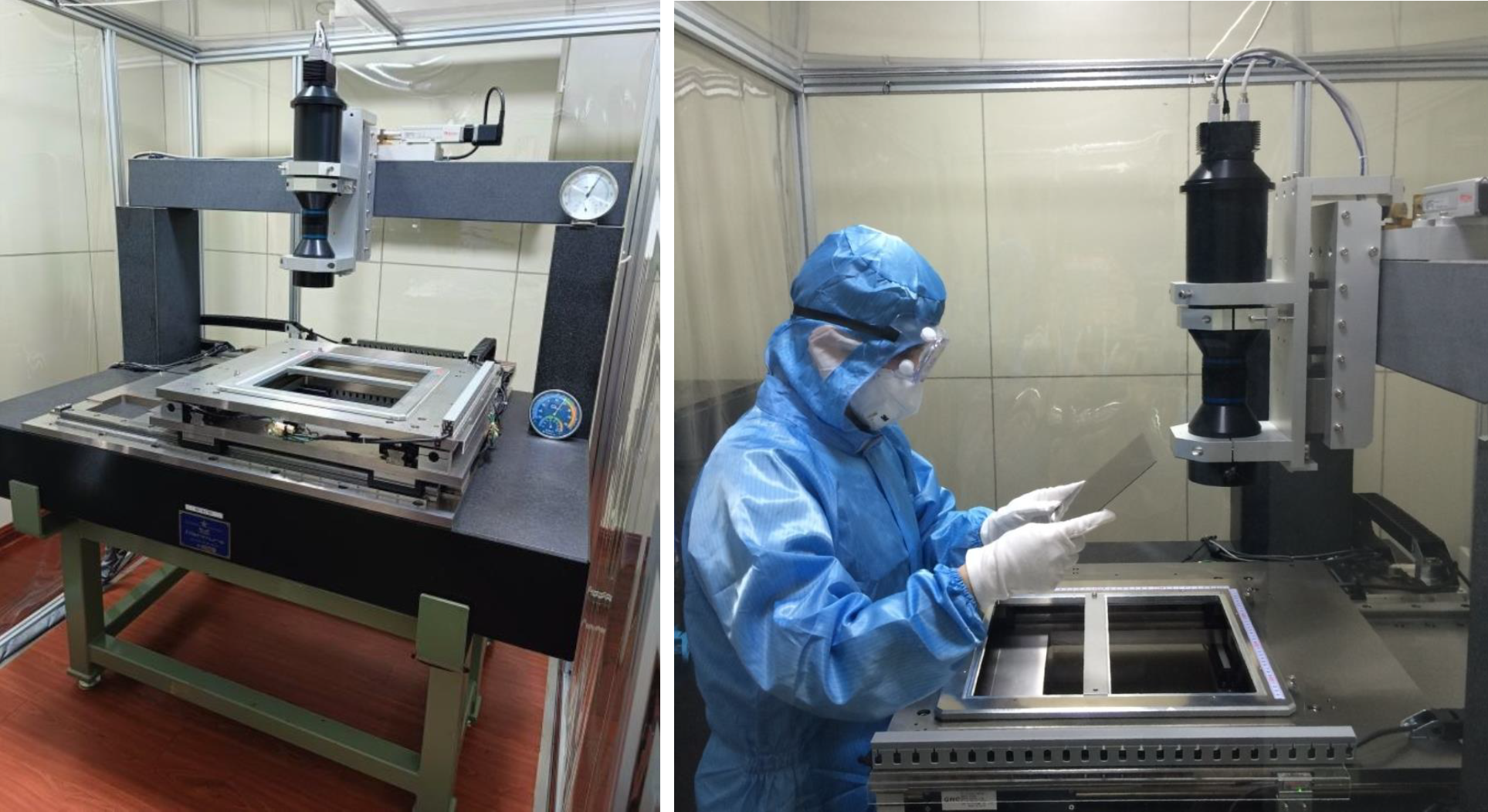}
  \caption{The first digitizer at SHAO.}
  \label{Fig2}
\end{figure}

Then, with the support from Chinese Ministry of Science and Technology and in collaboration with Nishimura Co. (Japan), SHAO developed its first dedicated plate digitizer (Fig.~\ref{Fig2}). The system, based on mechanical guide rails, included a telecentric lens, high-resolution camera, LED illumination, and a mobile platform mounted on granite.

The digitizer uses the line scanning method. For a large plate, after several strip scans (28 mm wide and less than 300 mm long), all the strip images are assembled to form a complete image. With this machine, we completed the digitization of ~30,000 Chinese astromomical plates.

Our astronomical plate digitization work was fully recognized by the Shanghai government, which authorized our laboratory to provide high-precision digitization services for other fields such as surveying and cartography, medicine, and biology.

\subsection{New Digitizers (2019--2024)}

\begin{figure}[!t]
  \includegraphics[width=\columnwidth]{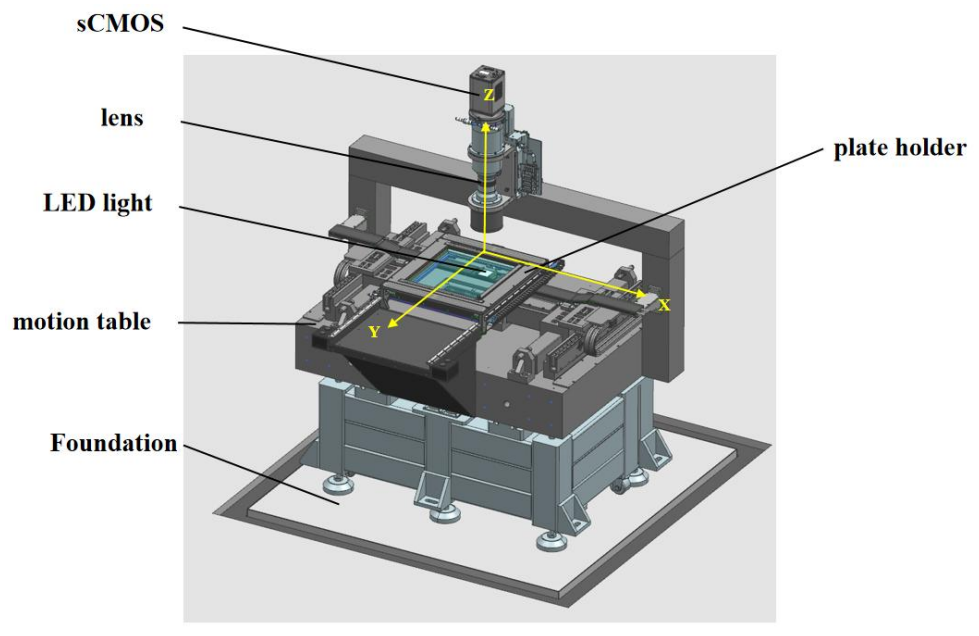}
  \caption{New generations of plate digitizers at SHAO.}
  \label{Fig3}
\end{figure}

In 2019, SHAO began developing a new digitizer based on air-bearing guide rails. Components included a high-precision sCMOS camera, air-bearing motion table, vibration-isolated foundation, electropneumatic plate holder, and fine-tuned camera alignment unit (Fig.~\ref{Fig3}). This new plate digitizer has a stable base and is built on a strong concrete isolated foundation to be isolated from vibrations and avoid vibrations from the outside.

Regarding the air-bearing motion table, we use a laser interferometer to check the accuracy and a marble square ruler to check the orthogonality.

After the table arrived, we carried out fine-tuning and performance tests to ensure that the table can meet the following requirements:
\begin{itemize}
\item Short-term positioning repeatability: $<$0.1 $\mu$m,
\item Straightness and flatness: $<$0.3$\mu$m,
\item Pitch/yaw: $<$1.0",
\item Orthogonality of 2 motion axes: $<$2.0".
\end{itemize}

Regarding the electro-pneumatic plate holder, we adopted the electric and pneumatic operation mode in order to:
\begin{itemize}
\item Load and unload plates of different sizes up to 320$\times$320 mm$^2$,
\item Keep the plate stable during the scanning process.
\end{itemize}

We also worked on fine-tuning the camera rotation, to align the camera's axes with those of the motion table. A fine-tuning mechanism was installed, allowing for precise camera rotation. The angular resolution of the adjustment is approximately 1.4 arc-seconds.

\begin{figure}[!t]
  \includegraphics[width=\columnwidth]{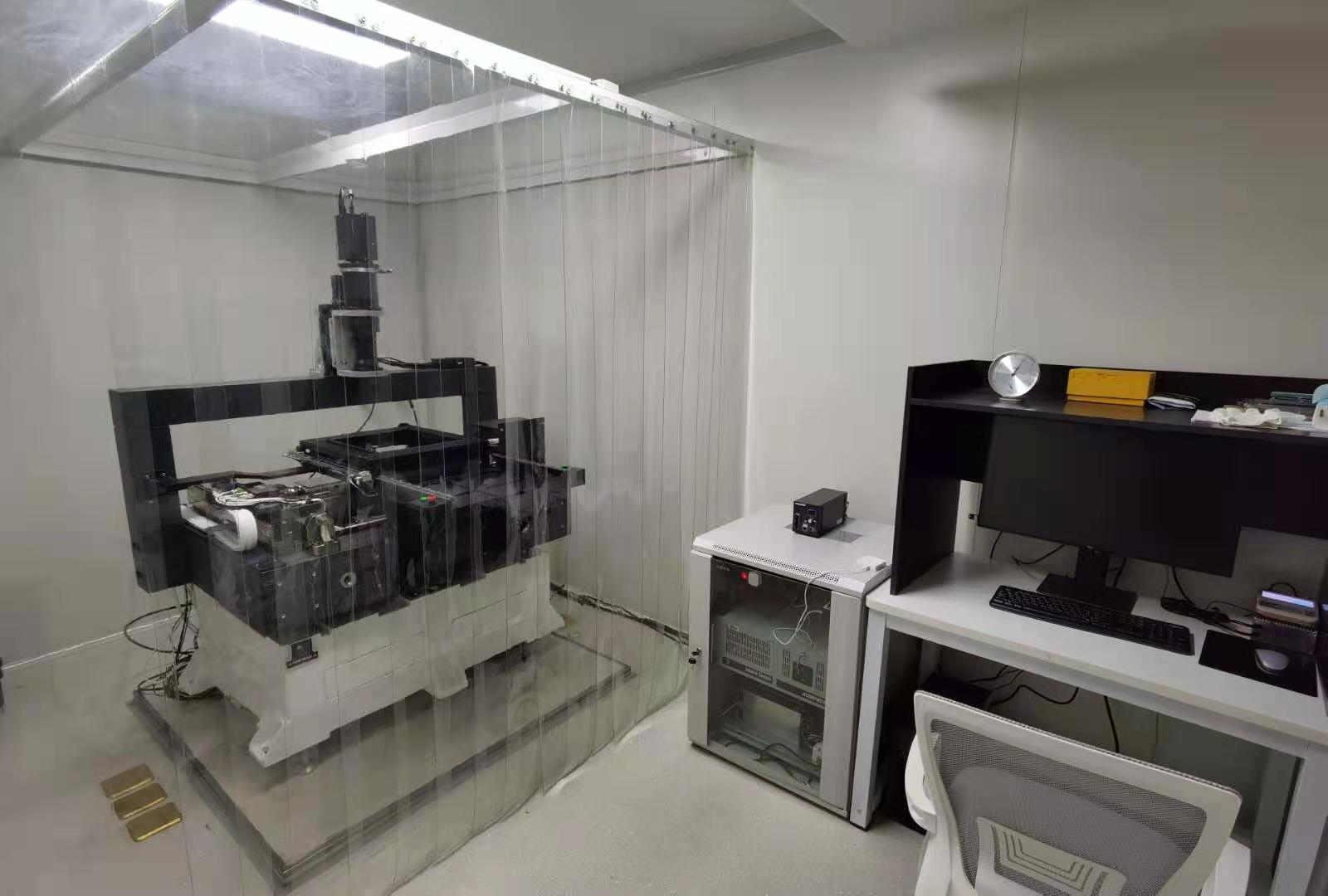}
  \caption{The first new digitizer at SHAO.}
  \label{Fig4}
\end{figure}

The development of the first new digitizer was completed at the end of 2020, as shown in the Fig.~\ref{Fig4}. Concerning the operation mode, this new digitizer uses a block scanning “step-and-stare” method. A 300$\times$300 mm$^2$ plate is digitized in ~210 seconds, generating ~1.45 GB of 16-bit FITS data with sub-micron positioning precision. An example of star images on a plate digitized by the new generation SHAO's machine is shown in the Fig.~\ref{Fig5} below.

\begin{figure}[!t]
  \includegraphics[width=\columnwidth]{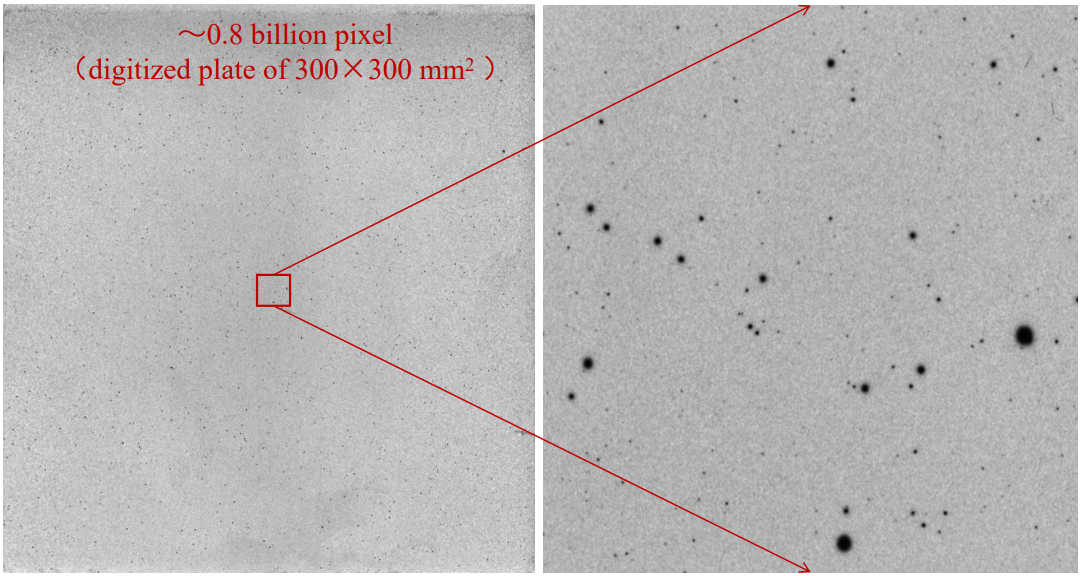}
  \caption{Example of star images on a plate digitized by the new generation SHAO's machine. At this digitized plate of 300$\times$300 mm$^2$ we have $\sim$0.8 billion pixels.}
  \label{Fig5}
\end{figure}

The main difference on parameters between the old and the new plate digitizers of SHAO are summarised in Tab.~\ref{Tab1}.

\begin{table}[ht]
\centering
\scriptsize 
\setlength{\tabcolsep}{3pt} 
\renewcommand{\arraystretch}{1.1} 
\caption{Main parameters of the old and the new generation plate digitizers at SHAO}
\label{Tab1}
\begin{tabular}{lcc 
                >{\raggedright\arraybackslash}p{3cm} 
               >{\raggedright\arraybackslash}p{3cm}}
\toprule
\textbf{Parameter} & \textbf{First Digitizer} & \textbf{New Digitizer}\\
\midrule
Scanning mode & Line scanning & Block scanning \\
Optical resolution & 2540 DPI & 2309 DPI \\
Positioning precision & 1.0$\mu$m & 0.3$\mu$m \\
Photometric precision & 0.02mag & 0.01mag \\
Dynamic range & 3.0 OD & 3.9 OD \\
Max scanning area & 300mm$\times$300mm & 320mm$\times$320mm \\
Scanning time & 620 s (300mm$^2$) & 210 s (300mm$^2$) \\
Image format & 16-bit FITS & 16-bit FITS \\
\bottomrule
\end{tabular}
\end{table}

\section{International Cooperations}

\begin{figure}[!t]
  \includegraphics[width=\columnwidth]{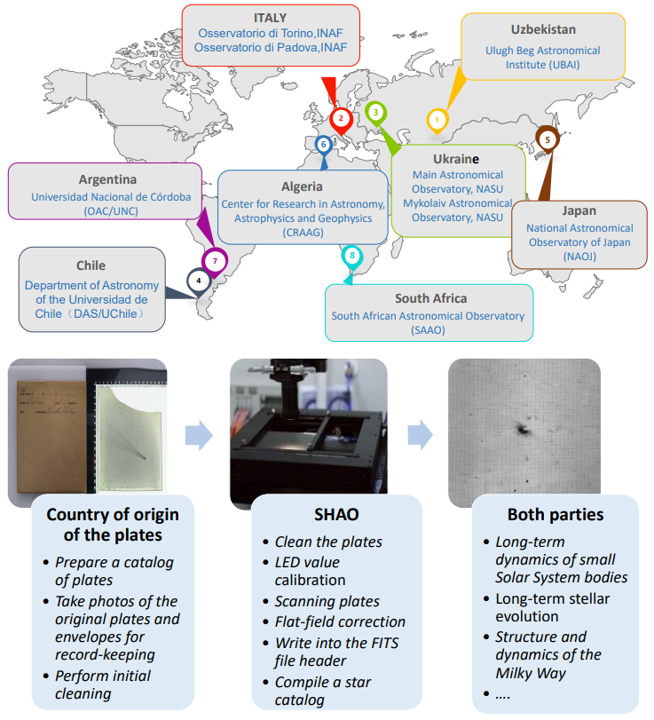}
  \caption{\textbf{Top:} International institutions currently engaged in collaborative projects. \textbf{Bottom:} Steps of bilateral collaboration with SHAO.}
  \label{Fig6}
\end{figure}

In response to the call of the IAU, SHAO has carried out international cooperation in the digitization of astronomical plates with multiple institutions using the high-precision digitizers.
Fig.~\ref{Fig6} depicts the main institutions currently engaged in collaborative projects as well as the steps of bilateral collaboration with SHAO.

\begin{figure}[!t]
  \includegraphics[width=\columnwidth]{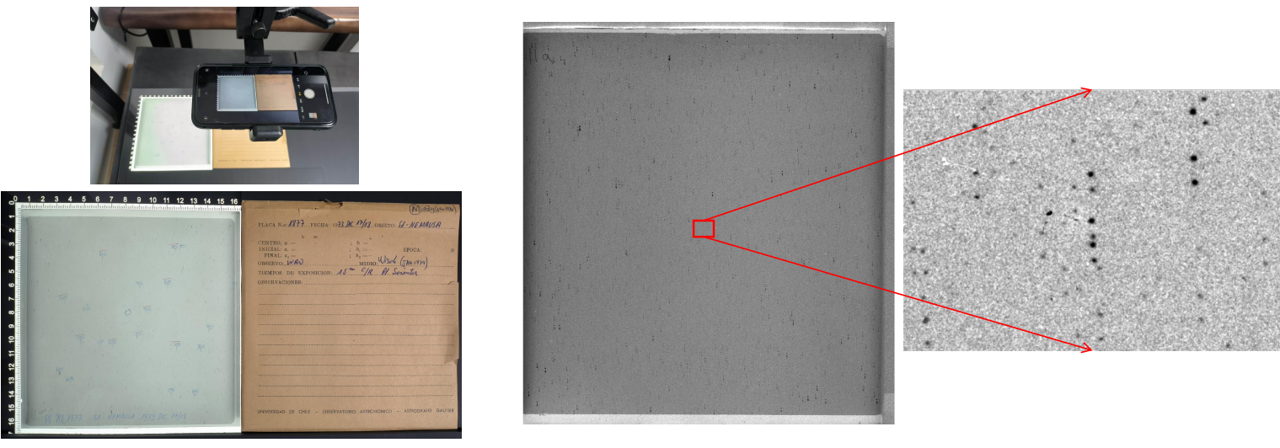}
  \caption{Recording the original information on the Chilean plates and envelopes (left), then digitization to obtain high-precision images from that plates (right).}
  \label{Fig7}
\end{figure}

A collaboration is also underway with the University of Chile, where 6,012 plates from its observatory have arrived at SHAO. Digitization and processing of the data from these plates are in progress.
We record the original information on the plates and envelopes, then use the new digitizer to obtain high-precision images from plates (Fig.~\ref{Fig7}).

\section{Scientific Application on Astronomical plates}
\begin{figure}[!t]
  \includegraphics[width=\columnwidth]{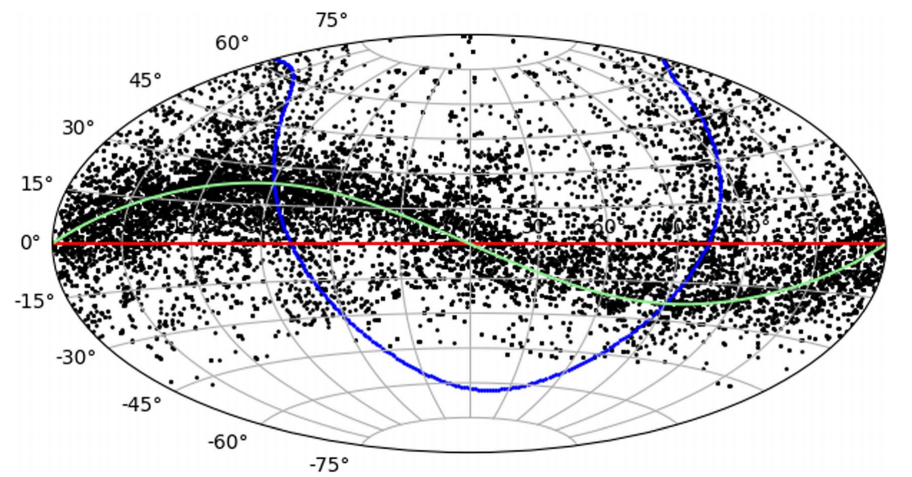}
  \caption{Distribution of the pointings of Chinese plates.}
  \label{Fig8}
\end{figure}

\begin{figure}[!t]
  \includegraphics[width=\columnwidth]{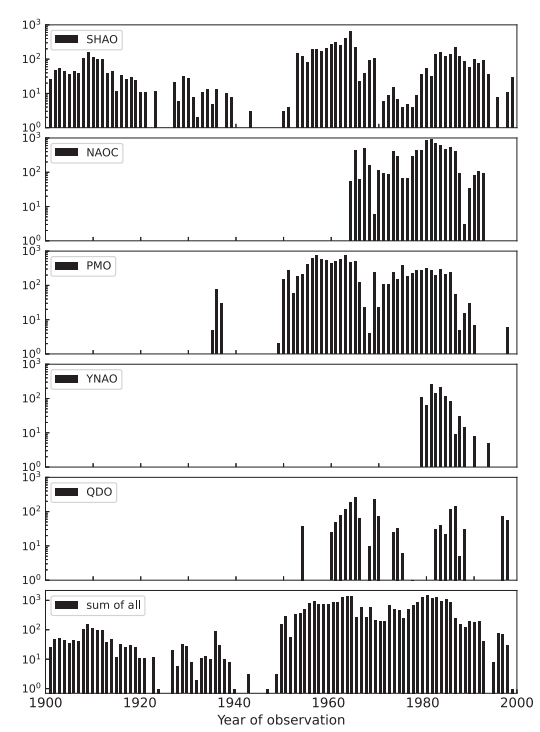}
  \caption{Distribution of the quantities of the astronomical plates from each station over time.}
  \label{Fig9}
\end{figure}

For Chinese astronomical plates, astrometric and photometric calibrations were conducted, with data made available through the China Virtual Observatory \citep[\url{https://nadc.china-vo.org/data/data/legacyplateedr/f};][]{Shang2024}.
This batch of data contains observations from 9 telescopes at five stations, namely, National Astronomical Observatory (NAOC), Shanghai Astronomical Observatory (SHAO), Purple Mountain Observatory (PMO), Yunnan Astronomical Observatory (YAO), and Qingdao Observatory (QDO), and spans the period from 1901 to 1998. Fig.~\ref{Fig8} illustrates the distribution of the pointings of the plates, and Fig.~\ref{Fig9} shows the distribution of the quantities of the plates from each station over time.

\begin{figure}[!t]
  \includegraphics[width=\columnwidth]{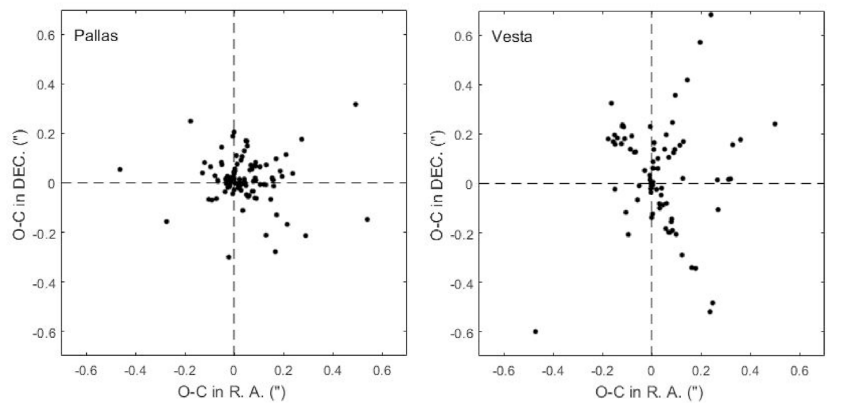}
  \caption{(O–C)s in R.A. versus (O–C)s in DEC.}
  \label{Fig10}
\end{figure}

In addition, Chinese and Uzbek researchers collaborated to carry out the new reduction of the plates of Pallas and Vesta. Base on Gaia DR2 catalog, we obtained the asteroids astrometric results, with an precision of 0.1” to 0.2” \citep{Wang2023}. Fig.~\ref{Fig10} shows the (O–C) in right ascension versus (O–C) in declination for Pallas and Vesta, where O represents the astrometric results from the plates reduction, and C represents the theoretical positions derived from the JPL Horizons System.

\section{Conclusion}

The plate digitization efforts at SHAO have advanced both technology and collaboration, preserving irreplaceable observational heritage while providing high-precision data for astronomical research.

\acknowledgements
This work is supported by International Partnership Program of Chinese Academy of Sciences (Grant No. 018GJHZ2023110GC) and National Natural Science Foundation of China (Grant No. 12473070).

\end{document}